\documentclass[final,5p,times,twocolumn]{elsarticle}

\usepackage{amssymb}
\usepackage{amsmath}
\usepackage{hyperref}
\usepackage{xcolor}

\usepackage{algorithm}
\usepackage{algpseudocode}
\usepackage{siunitx}
\usepackage[frozencache,cachedir=.]{minted}
\newlength{\mintednumbersep}
\AtBeginDocument{%
  \sbox0{\tiny00}%
  \setlength\mintednumbersep{\parindent}%
  \addtolength\mintednumbersep{-\wd0}%
}
\usepackage{subcaption}
\usepackage{caption}
\usepackage{float}
\usepackage{tcolorbox}
\usepackage{tikz}
\usepackage{booktabs}

\newenvironment{customlisting}{%
  \begin{figure}[t] 
  \captionsetup{type=listing} 
  \centering
  \begin{minipage}{\linewidth}
}{%
  \end{minipage}
  \end{figure}
}

\newenvironment{customlistingH}{%
  \begin{figure}[H] 
  \captionsetup{type=listing} 
  \centering
  \begin{minipage}{\linewidth}
}{%
  \end{minipage}
  \end{figure}
}

\biboptions{sort&compress}

\newcounter{bla}

\journal{Computer Physics Communications}

\definecolor{NavyBlue}{RGB}{0, 0, 128} 

\hypersetup{
  colorlinks=true,
}

\begin{document}
\hypersetup{
  linkcolor=NavyBlue,
  urlcolor=NavyBlue,
  citecolor=NavyBlue
}

\begin{frontmatter}



  \title{TorchOptics: An open-source Python library for differentiable Fourier optics simulations
  }


  \author{Matthew J. Filipovich\corref{author}}
  \author{A. I. Lvovsky}

  \cortext[author] {Corresponding author.\\\textit{E-mail address:} matthew.filipovich@physics.ox.ac.uk}
  \address{  Clarendon Laboratory, University of Oxford, Parks Road, Oxford, OX1 3PU, United Kingdom}

  \begin{abstract}
    TorchOptics is an open-source Python library for differentiable Fourier optics simulations, developed using PyTorch to enable GPU-accelerated tensor computations and automatic differentiation. 
    It provides a comprehensive framework for modeling, analyzing, and designing optical systems using Fourier optics, with applications in imaging, diffraction, holography, and signal processing. 
    The library leverages PyTorch's automatic differentiation engine for gradient-based optimization, enabling the inverse design of complex optical systems.
    TorchOptics supports end-to-end optimization of hybrid models that integrate optical systems with machine learning architectures for digital post-processing.    
    The library includes a wide range of optical elements and spatial profiles, and supports simulations with polarized light and fields with arbitrary spatial coherence.
  \end{abstract}


\end{frontmatter}



{\bf PROGRAM SUMMARY}

\begin{small}
  \noindent
  {\em Program Title:} TorchOptics                                          \\
  {\em Developer's repository link:} \\ \href{https://github.com/MatthewFilipovich/torchoptics}{https://github.com/MatthewFilipovich/torchoptics} \\
  {\em Licensing provisions:} MIT  \\
  {\em Programming language:} Python                         \\
  {\em Supplementary material:} Documentation is available at \href{https://torchoptics.readthedocs.io}{torchoptics.readthedocs.io}.
  The TorchOptics library and its dependencies can be installed from PyPI at \href{https://pypi.org/project/torchoptics}{pypi.org/project/torchoptics}.   \\
  {\em Nature of problem:} Differentiable Fourier optics simulations.  \\
  {\em Solution method:} TorchOptics uses numerical methods to simulate the evolution of optical fields in systems, integrating PyTorch's automatic differentiation engine to support gradient-based optimization. \\

\end{small}

\newpage
\section{Introduction}

Machine learning has driven significant advancements in the design and large-scale optimization of optical hardware, enabling the development of novel optical systems for applications in imaging, communication, and computing~\cite{lecunDeepLearning2015, linAllopticalMachineLearning2018, wetzsteinInferenceArtificialIntelligence2020, shastriPhotonicsArtificialIntelligence2021, gentyMachineLearningApplications2021,menguIntersectionOpticsDeep2022}. 
Furthermore, machine learning models, especially convolutional neural networks, are increasingly applied in the digital post-processing of optical signals. These models can improve system performance, such as achieving higher image resolution, without requiring modifications to the optical hardware~\cite{sinhaLenslessComputationalImaging2017, rivensonDeepLearningMicroscopy2017,dosterMachineLearningApproach2017, barbastathisUseDeepLearning2019}. 

Traditionally, free-space optical systems have been modeled using ray-tracing methods based on geometric optics approximations~\cite{smithModernOpticalEngineering2008}. However, these methods neglect wave optics phenomena, such as diffraction and interference, that can be leveraged for novel applications and enhanced performance~\cite{hughesWavePhysicsAnalog2019,zhouLargescaleNeuromorphicOptoelectronic2021, dorrahTunableStructuredLight2022}. Fourier optics presents an alternative framework for analyzing optical systems, where the behavior of light is treated using wave optics. In this approach, light propagation is modeled using scalar diffraction theory~\cite{goodmanIntroductionFourierOptics2017}. 

Computational Fourier optics is well-suited for the inverse design of free-space optical systems. Gradient-based optimization of complex optical components can be performed by incorporating automatic differentiation techniques into the simulations, enabling efficient gradient computations. This approach has been successfully applied in the design of diffractive optical neural networks, super-resolution microscopy setups, and metasurfaces~\cite{menguAnalysisDiffractiveOptical2020, rodriguezXLuminAAutodifferentiatingDiscovery2024c, hazinehDFlatDifferentiableFlatOptics2022}. 

Additionally, computational Fourier optics enables end-to-end optimization of hybrid models that integrate optical hardware with machine learning architectures for digital post-processing. 
This integrated approach jointly optimizes both components within a unified, fully differentiable model, using the backpropagation algorithm to efficiently calculate the loss gradients with respect to all model parameters~\cite{rumelhartLearningRepresentationsBackpropagating1986}. This approach has led to improved design solutions compared to separate optimization of hardware and software~\cite{sitzmannEndtoendOptimizationOptics2018, baekSingleshotHyperspectralDepthImaging2021, linEndtoendNanophotonicInverse2021}. 

In this paper, we introduce TorchOptics, an open-source Python library built using the PyTorch framework for differentiable Fourier optics simulations.
Motivated by the surge in research at the intersection of optics and machine learning, we developed TorchOptics as a robust, user-friendly simulation library for the modeling, analysis, and optimization of optical systems. 
TorchOptics harnesses the PyTorch framework to enable accelerated tensor computations using graphics processing units (GPUs) and automatic differentiation for efficient gradient calculations~\cite{paszkePyTorchImperativeStyle2019}.
The library was developed following standard PyTorch conventions for seamless integration with existing PyTorch workflows and features. 
The package can be installed directly from PyPI at \href{https://pypi.org/project/torchoptics/}{pypi.org/project/torchoptics}. Further documentation, including tutorials and the API reference, is available at \href{https://torchoptics.readthedocs.io}{torchoptics.readthedocs.io}.

TorchOptics simulations allow optical system properties, such as element positions and modulation profiles, to be treated as learnable parameters. These parameters are iteratively updated during training to minimize a specified objective function using gradient-based optimization algorithms, such as Adam or stochastic gradient descent~\cite{kingmaAdamMethodStochastic2017}. Gradients are computed using PyTorch's automatic differentiation engine, which constructs a dynamic graph of operations performed during the forward pass (i.e., the evolution of the input fields). During the backward pass, the backpropagation algorithm computes the gradients of the loss function with respect to the learnable parameters by traversing the graph and applying the rules of automatic differentiation. 

The library includes many standard optical elements, including lenses, polarizers, and waveplates. It also supports elements with custom modulation profiles, such as spatial light modulators (SLMs) and diffractive optical elements (DOEs). In addition to simulating scalar monochromatic light, TorchOptics supports polarized fields using Jones calculus, and fields with arbitrary spatial coherence through the mutual coherence function~\cite{rubinMatrixFourierOptics2019,goodmanStatisticalOptics2015}. Polychromatic light can be simulated by modeling each discrete wavelength in the optical spectrum separately, then incoherently summing their intensity distributions. The library also provides several wavefront profiles, including Hermite-Gaussian and Laguerre-Gaussian beams, as well as commonly used modulation profiles like diffraction gratings. 

This paper is organized as follows: In Sec.~\ref{sec:library}, we discuss the core functionality provided by the TorchOptics library and introduce the primary simulation classes. In Sec.~\ref{sec:field_operations}, we present the Fourier optics theory used to model the behavior of optical fields and describe its implementation in TorchOptics. Section~\ref{sec:gradient} discusses gradient-based optimization in TorchOptics simulations using PyTorch's automatic differentiation engine. Section~\ref{sec:additional_features} showcases additional features that extend simulation capabilities beyond coherent monochromatic light. Finally, we present our conclusions in Sec.~\ref{sec:conclusion}.

\section{TorchOptics overview}\label{sec:library}

TorchOptics is an object-oriented library for modeling optical fields and simulating their evolution through optical systems.
It consists of three core classes: \texttt{Field}, \texttt{Element}, and \texttt{System}. Each class includes methods for simulation and analysis, with properties stored as PyTorch tensors which can be dynamically updated with a single line of code. 
These classes inherit from PyTorch's \texttt{Module} class to enable standard PyTorch functionalities, including device management (CPU and GPU) and parameter registration for gradient-based optimization, which is further discussed in Sec.~\ref{sec:gradient}. 

\paragraph{Fields} Monochromatic optical fields are represented using the \texttt{Field} class, which encapsulates the complex-valued wavefronts sampled on a specified two-dimensional plane.
The class includes methods for modulation and propagation, which are further described in Sec.~\ref{sec:field_operations}. It also supports a range of methods for field analysis and manipulation, including visualization, normalization, and calculation of the intensity's centroid and standard deviation. 

\paragraph{Elements} Optical elements are implemented as subclasses of the \texttt{Element} class and are modeled as  planar surfaces in accordance with Fourier optics. TorchOptics includes several element types, such as modulators, detectors, and beam splitters. 
Each element applies a transformation to an input field defined in its \texttt{forward()} method. For instance, the \texttt{Lens} class, which models a thin lens, applies a quadratic phase factor to the field's wavefront.

\paragraph{Systems} The \texttt{System} class models an optical system as a sequence of \texttt{Element} objects arranged along the optical axis. 
Its \texttt{forward()} method calculates the evolution of an input field as it propagates through the system and returns the field after it has been processed by the final element. Additionally, the \texttt{measure()} method computes the field at any specified position within the system.\\

TorchOptics supports parallel processing of multiple optical fields by representing each field as an element along the tensor's batch dimension, similar to batch processing in standard PyTorch operations.
This technique can improve computational speed and efficiency during simulations and is particularly useful for training with mini-batches of input examples. Multi-GPU processing of TorchOptics simulations can be performed using PyTorch's data and model parallelism features. 

\begin{figure*}
  \centering
  \includegraphics[width=\linewidth]{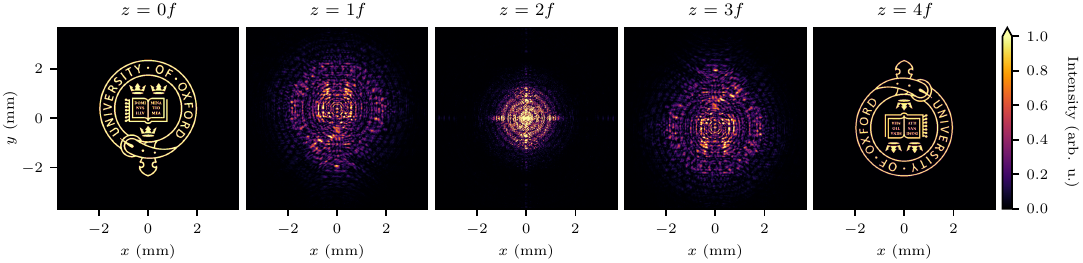}
  \caption{Intensity distributions of the optical field at each focal plane along the $z$-axis in a $4f$ imaging system. 
  The system consists of two lenses, each with a focal length of $f = \SI{200}{\milli\meter}$ and a diameter of \SI{10}{\milli\meter}. The lens profiles are simulated using 1000 grid points per dimension with \SI{10}{\micro\meter} spacing. The coherent input field has a uniform phase profile and a wavelength of $\lambda=\SI{700}{\nano\meter}$.}
  \label{fig:4f_system}
\end{figure*}

Listing~\ref{listing:4f_system} provides an example script for simulating a $4f$ imaging system using TorchOptics. The script defines the simulation properties and initializes both the input field and the optical system. The field is computed at each focal plane along the optical axis, and the corresponding intensity distributions are shown in Fig.~\ref{fig:4f_system}.

\begin{customlistingH}
\begin{minted}{python}
import torch
import torchoptics
from torchoptics import Field, System
from torchoptics.elements import Lens

# Set simulation properties
lens_shape = 1000  # Number of grid points in each dimension
focal_length = 200e-3  # Focal length of the lens (m)
spacing = 10e-6  # Spacing between grid points (m)
wavelength = 700e-9  # Wavelength of the light (m)

# Determine device
device = "cuda" if torch.cuda.is_available() else "cpu"

# Configure torchoptics default properties
torchoptics.set_default_spacing(spacing)
torchoptics.set_default_wavelength(wavelength)

# Initialize the input field with a loaded wavefront pattern
input_field = Field(torch.load("input_data.pt"), z=0).to(device)

# Define 4f optical system with two lenses
system = System(
    Lens(lens_shape, focal_length, z=1 * focal_length),
    Lens(lens_shape, focal_length, z=3 * focal_length),
).to(device)

# Measure field at focal planes along the z-axis
measurements = [
    system.measure_at_z(input_field, z=i * focal_length) 
    for i in range(5)
]

# Visualize the measured fields
for measurement in measurements:
    measurement.visualize()
\end{minted}
  \caption{Python script for simulating a $4f$ imaging system using TorchOptics. The input field is initialized using a spatial profile loaded from the \texttt{input\_data.pt} file.  The field at each focal plane along the $z$-axis is calculated and visualized. The corresponding intensity distributions are shown in Fig.~\ref{fig:4f_system}.  
  }
  \label{listing:4f_system}
\end{customlistingH}

\section{Computational Fourier optics}\label{sec:field_operations}

Fourier optics, founded on the principles of Fourier analysis, provides a powerful approach for analyzing the behavior of light within optical systems~\cite{goodmanIntroductionFourierOptics2017}. In this framework, monochromatic optical fields are described on two-dimensional planes along the $xy$-axes, with the scalar field distribution at a given $z$-plane represented as $\psi_z(x, y)$. TorchOptics models this planar field as a discrete grid that is uniformly sampled over the $xy$-plane. The grid is initialized with a specified size and spacing interval along the $x$- and $y$-axes, and can be offset relative to the $z$-axis. 

The evolution of optical fields in systems is described by two fundamental operations: modulation by optical elements and free-space propagation. 
TorchOptics simulates these operations using methods implemented in the \texttt{Field} class, which encapsulates the sampled field distribution as a PyTorch tensor. 
The \texttt{System} class executes these modulation and propagation methods in the correct sequence to compute the field's evolution through the optical system.

Modulation of an input optical field $\psi_\mathrm{in}(x, y)$ by a complex-valued modulation profile $\mathcal{M}(x, y)$ is realized as a point-wise product. The resulting output field $\psi_\mathrm{out}(x, y)$ is expressed as
\begin{equation}\label{eq:field_modulation}
  \psi_\mathrm{out}(x, y) = \mathcal{M}(x, y) \, \psi_\mathrm{in}(x, y).
\end{equation}

\begin{figure}[t]
  \centering
  \includegraphics[width=\linewidth]{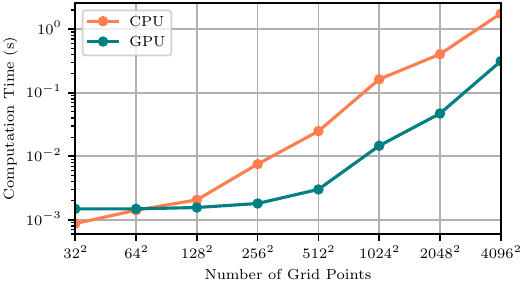}
  \caption{Computation time of field propagation in TorchOptics as a function of grid size. The evaulation was performed on both CPU (i9-13900K Processor) and GPU (RTX 5000 Ada Generation) using the DI method with double-precision floating-point values.}
  \label{fig:runtime}
\end{figure}

Free-space propagation of optical fields between planes is treated using scalar diffraction theory. 
The Rayleigh-Sommerfeld diffraction integral describes the propagation of an input field $\psi_0(x', y')$ at $z=0$ to the position ($x, y, z$) as
\begin{equation}\label{eq:field_propagation}
  \psi_z(x, y) = \iint h_{z}(x-x', y-y') \, \psi_0(x', y') \, \mathrm{d}x' \mathrm{d}y'.
\end{equation}
Here, the impulse response $h_z$ is given by
\begin{equation}\label{eq:impulse_response}
  h_z(x-x', y-y') = \frac{z}{i\lambda d} \left(1 + \frac{i}{kd}  \right) \frac{\exp \left({i k d} \right)}{d},
\end{equation}
where  $d=\sqrt{(x-x')^2+(y-y')^2+z^2}$ is the Euclidean distance between the input and output positions, $\lambda$ is the wavelength, and $k=2\pi / \lambda$ is the wavenumber~\cite{goodmanIntroductionFourierOptics2017}. TorchOptics numerically evaluates the diffraction integral in Eq.~\eqref{eq:field_propagation} using the fast Fourier transform (FFT) algorithm by treating it as a convolution operation. This approach for modeling field propagation is known as the \textit{Direct Integration} (DI) method~\cite{shenFastFouriertransformBasedNumerical2006, voelzComputationalFourierOptics2011}. Figure~\ref{fig:runtime} shows the computation time for field propagation with different field sizes on both CPU and GPU.

\begin{figure}[t]
  \centering
  \includegraphics[width=\linewidth]{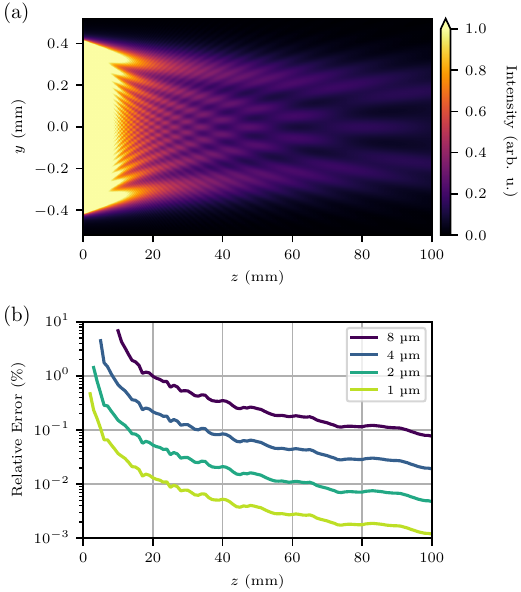}
  \caption{
    Field propagation and error analysis for a one-dimensional aperture simulation. 
    (a)~Intensity distribution of the field along the $yz$-plane with wavelength $\lambda=\SI{700}{\nano\meter}$. The aperture at $z=\SI{0}{\milli\meter}$ has a length of \SI{0.8}{\milli\meter} along the $y$-axis.
    (b)~Relative error of the intensity distribution along the $z$-axis using the DI propagation method, compared to the solution determined using the \texttt{quad} numerical integration method from the SciPy package~\cite{virtanenSciPyFundamentalAlgorithms2020}. 
    The error is shown for grid spacing intervals of 1, 2, 4, and \SI{8}{\micro\meter} and represents the average relative error of intensity values spaced \SI{8}{\micro\meter} apart, for $|y|\leq\SI{0.4}{\milli\meter}$ and $z$-values satisfying the sampling condition in Eq.~\eqref{eq:sampling_condition}.
    }
  \label{fig:accuracy}
\end{figure}

Alternatively, field propagation can be analyzed in the spatial-frequency domain:
\begin{equation}\label{eq:field_propagation_tf}
  \psi_z(x, y) = \mathcal{F}^{-1} \left\{ H_z\left(k_x, k_y\right) \, \mathcal{F} \left\{ \psi_0\left(x', y'\right) \right\}  \right\},
\end{equation}
where $\mathcal{F}$ and $\mathcal{F}^{-1}$  are the 2D Fourier transform and its inverse across the $x$- and $y$-axes. The transfer function $H_z$ is 
\begin{equation}\label{eq:transfer_function}
  H_z(k_x, k_y) = \exp \left( i z \sqrt{k^2 - k_x^2 - k_y^2} \, \right).
\end{equation}
In this approach, the field is treated as a superposition of plane waves traveling in different directions, each acquiring a phase delay during propagation. TorchOptics computes the propagated field in Eq.~\eqref{eq:field_propagation_tf} using the FFT algorithm, with configurable zero-padding to reduce numerical artifacts. This computational approach is commonly referred to as the \textit{Angular Spectrum} (AS) method~\cite{shenFastFouriertransformBasedNumerical2006, voelzComputationalFourierOptics2011}.

TorchOptics implements both propagation approaches as fully differentiable functions. By using the FFT algorithm, the computational complexity of both methods is $\mathcal{O}(MN\log(MN))$ for an $M\times N$ sampled field. Additionally, TorchOptics supports solutions using the Fresnel diffraction integral, which is an approximation of the Rayleigh-Sommerfeld diffraction integral. Further implementation details of these numerical methods are given in Ref.~\cite{shenFastFouriertransformBasedNumerical2006}. 

Both the DI and AS methods introduce discretization errors from sampling continuous functions. 
The DI method has better sampling accuracy at larger propagation distances due to the complex exponential term in the impulse response $h_z$, whose phase variation decreases as $z$ increases. 
Conversely, the AS method achieves better accuracy at shorter distances, as the phase variation of the complex exponential term in the transfer function $H_z$ reduces with decreasing $z$. A comprehensive discussion of the sampling criteria for different sampling regimes is provided in Ref.~\cite{voelzComputationalFourierOptics2011}, which defines the following critical propagation distance:
\begin{equation}\label{eq:sampling_condition}
  z_{\mathrm{critical}} = \frac{L \, \Delta x}{\lambda},
\end{equation}
where $L$ is the aperture length and $\Delta x$ is the grid spacing. The AS method is preferred for propagation distances ${z<z_{\mathrm{critical}}}$; otherwise, the DI method is more suitable. By default, TorchOptics selects the appropriate method based on this criterion.   

Proper sampling of the continuous optical field requires choosing sampling sizes and spacing intervals that adequately capture its features. Selecting appropriate spacing intervals is particularly important for minimizing numerical artifacts, such as aliasing, and affects the accuracy of the propagation solution. The numerical error in the intensity distributions of propagated fields with different spacing intervals is illustrated in Fig.~\ref{fig:accuracy}. 

The propagation methods require that the input and output sampling grids have the same spacing intervals. 
If the spacings vary, TorchOptics first propagates the field to an intermediate grid with the same spacing as the input grid, followed by interpolation onto the output grid.
The interpolation method, such as bilinear or nearest-neighbor, can be specified. This approach is useful for simulating propagation between elements with different geometries, such as modulators with varying pixel pitches. 

\begin{customlisting}
\begin{minted}{python}
import torch
import torchoptics
from torchoptics import Field, Param, System
from torchoptics.elements import PhaseModulator
from torchoptics.profiles import gaussian

# Set simulation properties
shape = 250  # Number of grid points in each dimension
waist_radius = 150e-6  # Waist radius of the Gaussian beam (m)
spacing = 10e-6  # Spacing between grid points (m)
wavelength = 700e-9  # Wavelength of the light (m)

# Determine device
device = "cuda" if torch.cuda.is_available() else "cpu"

# Configure torchoptics default properties
torchoptics.set_default_spacing(spacing)
torchoptics.set_default_wavelength(wavelength)

# Initialize input and target fields
input_field = Field(gaussian(shape, waist_radius), z=0).to(device)
target_field = Field(torch.load("target.pt"), z=0.6).to(device)

# Define optical system with three trainable phase modulators
elements = [
    PhaseModulator(Param(torch.zeros(shape, shape)), z)
    for z in [0.0, 0.2, 0.4]
]
system = System(*elements).to(device)

# Train system to match the output and target fields
optimizer = torch.optim.Adam(system.parameters(), lr=0.1)
for iteration in range(400):
    optimizer.zero_grad()
    output_field = system.measure_at_z(input_field, 0.6)
    loss = 1 - output_field.inner(target_field).abs().square()
    loss.backward()
    optimizer.step()
\end{minted}
  \caption{Python script for training an optical system to split an input Gaussian beam into four separate Gaussian beams. The target field is initialized from a spatial profile loaded from the \texttt{target.pt} file. The system consists of three modulators with trainable phase profiles. During training, the difference between the output and target fields is minimized over 400 iterations using the Adam optimizer. The training results are shown in Fig.~\ref{fig:gaussian_splitter}.}
  \label{listing:gaussian_splitter}
\end{customlisting}

\begin{figure}[t]
  \centering
  \includegraphics[width=\linewidth]{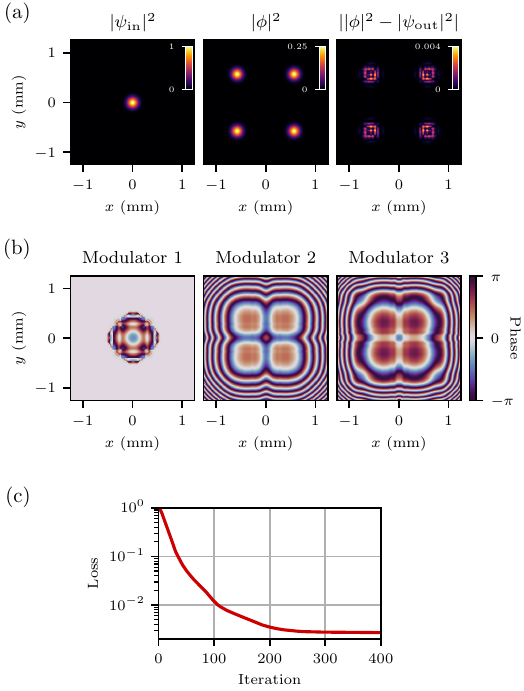}
  \caption{
    Training results of an optical system consisting of three phase modulators optimized to split a Gaussian beam. 
    (a) Intensity distributions of the input Gaussian field $\psi_\mathrm{in}$ and target field $\phi$, and the absolute difference between the intensity distributions of the target field and output field $\psi_\mathrm{out}$. (b) Trained phase modulation profiles of the three modulators. (c) Loss at each iteration during training.
  }
  \label{fig:gaussian_splitter}
\end{figure}

\section{Gradient-based optimization}\label{sec:gradient}
TorchOptics was developed for seamless integration with PyTorch's powerful gradient-based optimization framework.
Optical systems simulated with TorchOptics can be optimized using the same training workflow as standard PyTorch models, leveraging the machine learning optimization algorithms available in the \texttt{torch.optim} subpackage.

All continuous-valued properties of optical fields and elements can be registered as trainable PyTorch \texttt{Parameter} tensors. 
A property is designated as trainable during initialization by wrapping it with the TorchOptics \texttt{Param} class. 
PyTorch's automatic differentiation engine tracks these trainable properties and calculates their gradients during backpropagation, which are subsequently used by the optimizer to update the parameters.

TorchOptics enables end-to-end optimization of hybrid models that combine optical systems and machine learning architectures. PyTorch's automatic differentiation engine ensures that gradients flow throughout the entire model, enabling simultaneous optimization of all parameters in a single pass. This approach can be used to enhance the performance of an optical imaging system where the output signals are fed into a neural network for post-processing, optimizing both the optical and digital parameters in a holistic manner. 

During optimization, constraints on the parameters of optical systems are often required.
For instance, the amplitude attenuation applied by modulators must remain within the range of zero to one.
Constraints can be imposed by applying a transformation function to the parameter using the PyTorch subpackage \texttt{torch.nn.utils.parametrize}. 
The transformation is automatically applied before the tensor is used, ensuring the constraint is always satisfied throughout the training procedure.

A common optimization problem in optical system design involves training the modulation profiles of modulator elements (e.g., SLMs) to achieve a specified objective. Listing~\ref{listing:gaussian_splitter} presents a script for training an optical system with three trainable phase modulators to split a normalized incident Gaussian beam $\psi_{\mathrm{in}}(x, y)$ into four separate beams. The specified loss function $\mathcal{L}$ is proportional to the negative squared magnitude of the inner product between the output field $\psi_{\mathrm{out}}(x, y)$ and the target field $\phi(x, y)$:  
\begin{equation}\label{eq:loss_func}
  \mathcal{L} = 1 - \left| \, \iint \psi_{\mathrm{out}}(x, y) \, \phi^*(x, y) \, \mathrm{d}x \, \mathrm{d}y \, \right|^2
\end{equation}
The optimization results are shown in Fig.~\ref{fig:gaussian_splitter}.

\section{Advanced features}\label{sec:additional_features}
In this section, we introduce additional features that extend the simulation capabilities of TorchOptics beyond scalar, monochromatic fields. 
These features include the simulation of polarized fields, fields with arbitrary spatial coherence, and polychromatic fields.
Additionally, we present functions for generating spatial profiles commonly used in optics simulations.

\begin{customlisting}
\begin{minted}{python}
import torch
import torchoptics
from torchoptics import PolarizedField
from torchoptics.elements import LinearPolarizer

# Set simulation properties
shape = 1000
torchoptics.set_default_spacing(10e-6)
torchoptics.set_default_wavelength(700e-9)

# Initialize polarized field
field_data = torch.zeros(3, shape, shape)
field_data[0] = 1  # Polarized along x-axis
field = PolarizedField(field_data).normalize()

# Initialize linear polarizers
polarizer_45 = LinearPolarizer(shape, theta=torch.pi / 4)
polarizer_90 = LinearPolarizer(shape, theta=torch.pi / 2)

# Print the total power after transformations
print(field.power())  # 1
print(polarizer_45(field).power())  # 0.5
print(polarizer_90(field).power())  # 0
print(polarizer_90(polarizer_45(field)).power())  # 0.25
\end{minted}
  \caption{
    Python script for simulating the interaction between polarized fields and linear polarizers. A normalized input field polarized along the $x$-axis is defined. Two linear polarizers are initialized at \SI{45}{\degree} ($\pi/4$) and \SI{90}{\degree} ($\pi/2$) relative to the $x$-axis. The total power of the field is calculated and printed after passing through each polarizer individually and through both in sequence.
    }
  \label{listing:polarizers}
\end{customlisting}

\subsection{Polarization}
Polarization is a fundamental property of electromagnetic waves that affects their behavior in optical systems. 
In TorchOptics, polarization is modeled using matrix Fourier optics, an extension of Fourier optics that enables the simulation of fields with spatially varying polarization and their interactions with polarization-dependent optical elements~\cite{rubinMatrixFourierOptics2019}. 

Matrix Fourier optics uses Jones calculus to represent polarized optical fields as Jones vectors~\cite{hechtOptics2002}. A polarized field at the spatial position $\vec r$ can be expressed in Dirac notation as
\begin{equation}
  | \psi(\vec r) \rangle = \begin{pmatrix}
    \psi_H(\vec r) \\
    \psi_V(\vec r)
  \end{pmatrix},
\end{equation}
where $\psi_H(\vec r)$ and $\psi_V(\vec r)$ are the horizontal and vertical components of the field, respectively, perpendicular to the propagation direction.

Similarly, the transformations applied to polarized fields by polarization-dependent optical elements are represented by spatially varying Jones matrices. 
Each Jones matrix describes the operation performed on the field's horizontal and vertical polarization components at the position $\vec r$ and is expressed as
\begin{equation}
  \hat J (\vec r) = \begin{pmatrix}
    J_{HH}(\vec r) & J_{HV}(\vec r) \\
    J_{VH}(\vec r) & J_{VV}(\vec r)
  \end{pmatrix}.
\end{equation}
The output polarized field $|\psi_\mathrm{out}(\vec r)\rangle$ is given by the product of the Jones matrix and the input Jones vector: 
\begin{equation}
  |\psi_\mathrm{out}(\vec r)\rangle = \hat J(\vec r) \, |\psi_\mathrm{in}(\vec r)\rangle.
\end{equation}

The propagation of polarized fields is modeled by independently propagating each of its components, as polarized states are conserved during free-space propagation. 

TorchOptics models polarized fields using the \texttt{PolarizedField} class, a subclass of \texttt{Field}. 
This class represents the three-dimensional components of a polarized field, including $z$-polarization, as the propagation direction is not restricted solely to the $z$-axis.

TorchOptics includes several standard polarization-controlling optical elements, such as linear and circular polarizers, and arbitrary waveplates. Additionally, it supports custom elements that apply spatially varying polarization transformations. Listing~\ref{listing:polarizers} provides a simulation script demonstrating the interaction between polarized fields and linear polarizers.

\begin{figure}[t]
  \centering
  \includegraphics[width=\linewidth]{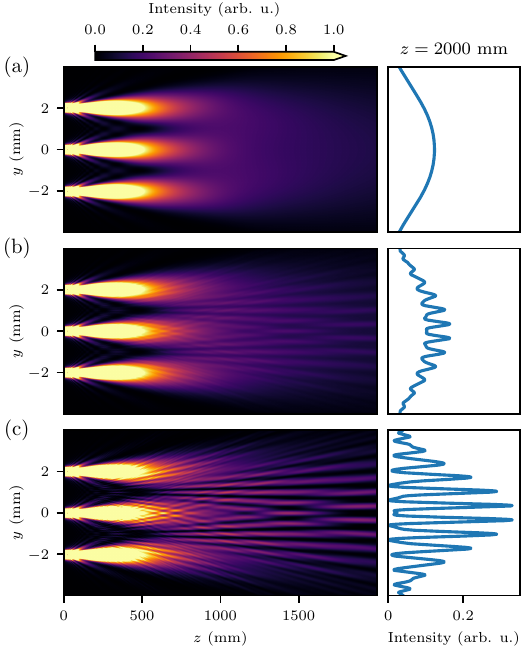}
  \caption{
    Diffraction patterns from three circular apertures with varying spatial coherence. The apertures are separated by ${\SI{2}{\milli\meter}}$ and each has a radius of \SI{200}{\micro\meter}. The intensity distributions along the $yz$-plane at ${x=\SI{0}{\milli\meter}}$ (left) and corresponding intensity profiles at ${z=\SI{2000}{\milli\meter}}$ (right) are illustrated for wavelength $\lambda=\SI{700}{\nano\meter}$.
    Three different spatial coherence conditions are shown: (a)~Incoherent light; (b)~Partially coherent light; and (c)~Fully coherent light.
  }
  \label{fig:spatial_coherence}
\end{figure}

\begin{customlisting}
\begin{minted}{python}
import torch
import torchoptics
from torchoptics import CoherenceField
from torchoptics.profiles import gaussian_schell_model

# Set simulation properties
shape = 50  
waist_radius = 50e-6  # Field waist radius
coherence_width = 25e-6  # Spatial coherence width
torchoptics.set_default_spacing(10e-6)
torchoptics.set_default_wavelength(700e-9)

# Initialize field spatial coherence with Gaussian Schell model 
data = gaussian_schell_model(shape, waist_radius, coherence_width)
coherence_field = CoherenceField(data)

# Propagate the coherence field to different planes along z-axis
propagated_coherence_fields = [
    coherence_field.propagate_to_z(z)
    for z in [0, 0.02, 0.04]
]
\end{minted}
  \caption{
    Python script for simulating the propagation of partially coherent light with a Gaussian Schell model distribution. 
  }
  \label{listing:coherence}
\end{customlisting}

\subsection{Spatial coherence}
Spatial coherence, which measures the correlation of the field's phase at different positions, plays an important role in the performance of many optical systems, notably in imaging and information processing applications~\cite{goodmanStatisticalOptics2015, filipovichRoleSpatialCoherence2024}. Figure~\ref{fig:spatial_coherence} demonstrates the effect of spatial coherence through diffraction patterns generated by incoherent, partially coherent, and fully coherent light. 

The spatial coherence of optical fields is characterized by the mutual coherence function, which determines the time-averaged correlation of the field at two spatial positions, ${\vec r_1=(x_1, y_1)}$ and ${\vec r_2=(x_2, y_2)}$, in the same plane~\cite{mandelOpticalCoherenceQuantum1995}. This function is defined as
\begin{equation}
  \Gamma_z(\vec r_1, \vec r_2)= \left \langle  \psi_z(\vec r_1; t) \, \psi_z^*(\vec r_2;t) \right\rangle_t,
\end{equation}
where $\left\langle \cdot \right\rangle_t$ is the time average. The time-averaged intensity of the field at the position $\vec r$ is given by $\Gamma (\vec r, \vec r)$.

The transformation applied by an optical element with modulation profile $\mathcal{M}(\vec r)$ to an input field with mutual coherence function $\Gamma_\mathrm{in}(\vec r_1, \vec r_2)$ is described by
\begin{equation}
  \Gamma_\mathrm{out}(\vec r_1, \vec r_2) = \mathcal{M}(\vec r_1) \, \mathcal{M}^*(\vec r_2) \, \Gamma_\mathrm{in}(\vec r_1, \vec r_2),
\end{equation}
where $\Gamma_\mathrm{out}(\vec r_1, \vec r_2)$ is the mutual coherence function of the output field.

Similar to Eq.~\eqref{eq:field_propagation}, the propagation of a field with mutual coherence function $\Gamma_0(\vec r_1\hspace{-0.3em}', \vec r_2\hspace{-0.3em}')$ at $z=0$  is given by
\begin{multline}
  \Gamma_z(\vec r_1, \vec r_2) = \iiiint \Big( h_z(\vec r_1 - \vec{r}_1\hspace{-0.3em}') \, h_z^*(\vec r_2 - \vec{r}_2\hspace{-0.3em}') \\
  \times \Gamma_0(\vec{r}_1\hspace{-0.3em}', \vec{r}_2\hspace{-0.3em}') \, \mathrm{d}\vec{r}_1\hspace{-0.3em}' \, \mathrm{d}\vec{r}_2\hspace{-0.3em}'  \Big),
\end{multline}
where the impulse response $h_z$ is defined in Eq.~\eqref{eq:impulse_response}.
The propagation can also be expressed in the spatial-frequency domain, analogous to Eqs.~\eqref{eq:field_propagation_tf} and \eqref{eq:transfer_function}.

In TorchOptics, optical fields with arbitrary spatial coherence are represented using the \texttt{CoherenceField} class, a subclass of \texttt{Field}. This class encapsulates the mutual coherence function as a four-dimensional tensor, enabling optics simulations with arbitrary spatial coherence. However, the use of the mutual coherence function significantly increases both memory and computational requirements, scaling quadratically relative to the \texttt{Field} class~\cite{filipovichRoleSpatialCoherence2024}. 

Listing~\ref{listing:coherence} presents a script for simulating the propagation of an optical field defined by the Gaussian Schell model, where both the intensity and coherence profiles follow Gaussian distributions. This model is often used to investigate the role of partially coherent light in optical systems~\cite{goodmanStatisticalOptics2015}. 

\subsection{Polychromatic fields}
In many practical scenarios, optical fields are not monochromatic but consist of a spectrum of optical wavelengths $\psi_z(\vec r; \lambda)$. 
These wavelength components are incoherent with respect to each other, such that the total intensity of the field $I_z(\vec r)$ is obtained by integrating over the intensities of the individual components:
\begin{equation}\label{eq:I_poly}
  I_z(\vec r) = \int \left|\psi_z(\vec r; \lambda)\right|^2 \, \mathrm{d}\lambda.
\end{equation}

\begin{figure*}[t]
  \centering
  \includegraphics[width=\linewidth]{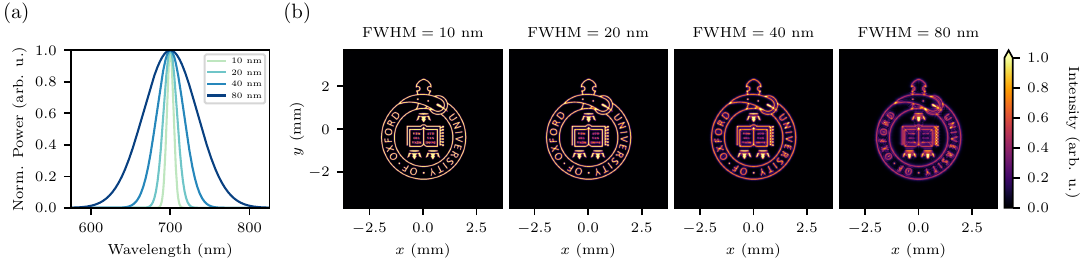}
  \caption{Imaging fields with Gaussian-shaped optical spectra using a $4f$ system. (a) Gaussian-shaped optical spectra with full-width at half-maximum (FWHM) values of 10, 20, 40, and \SI{80}{\nano\meter}. (b) Corresponding output intensity distributions from the $4f$ system for each spectrum. The optical system has the same configuration as described in Fig.~\ref{fig:4f_system}. 
  }
  \label{fig:bandwidth}
\end{figure*}

In TorchOptics, polychromatic fields are modeled by discretizing the optical spectrum and treating each wavelength as an individual monochromatic field during simulation. The total intensity across the spectrum is then calculated using Eq.~\eqref{eq:I_poly}. Listing~\ref{listing:bandwidth} demonstrates an example script for modeling polychromatic light using TorchOptics. The simulated chromatic aberrations caused by polychromatic light in a $4f$ imaging system are illustrated in Fig.~\ref{fig:bandwidth}.

\begin{customlisting}
\begin{minted}{python}
# ... (Code to define the system and gaussian_weighting function)
input_data = torch.load("input_data.pt")
wavelengths = torch.linspace(575e-9, 825e-9, 251)
weights = gaussian_weighting(wavelengths, mean=700e-9, fwhm=10e-9)
total_output_intensity = torch.zeros_like(input_data)

for wavelength, weight in zip(wavelengths, weights):
    input_field = Field(input_data * weight, wavelength=wavelength)
    output_field = system(input_field)
    total_output_intensity += output_field.intensity()

print(total_output_intensity)
\end{minted}
  \caption{
    Python script for simulating the evolution of polychromatic light with a Gaussian-shaped optical spectrum through a system. 
    Optical fields with wavelengths ranging from \SI{575}{\nano\meter} to \SI{825}{\nano\meter} are individually propagated through the system, each weighted according to the Gaussian spectral distribution. The total output intensity is computed as the incoherent sum of intensity contributions from all wavelengths. The intensity distributions for varying optical spectra propagated through a $4f$ system are shown in Fig.~\ref{fig:bandwidth}.}
  \label{listing:bandwidth}
\end{customlisting}

\subsection{Profiles}
The \texttt{torchoptics.profiles} subpackage provides a set of spatial profiles commonly used in optical simulations, including:
\begin{itemize}
\item \textbf{Beam profiles}: Gaussian, Hermite-Gaussian, and Laguerre-Gaussian beams.
\item \textbf{Modulation profiles}: Phase and amplitude diffraction gratings, and thin lenses.
\item \textbf{Aperture profiles}: Rectangular, circular, and checkerboard patterns. 
\item \textbf{Mutual coherence functions}: Schell Gaussian model and Schell model with arbitrary profiles. 
\end{itemize}

Profiles are generated by specifying the planar properties of the grid, including size, spacing intervals, and offsets along the $x$- and $y$-axes. Listings~\ref{listing:gaussian_splitter} and~\ref{listing:coherence} demonstrate the initialization of a Gaussian beam profile and Gaussian Schell model, respectively. 

\section{Conclusion}\label{sec:conclusion}

TorchOptics is an open-source Python library developed to address the growing interest in optical system design through machine learning techniques. It provides a powerful object-oriented framework for the simulation, analysis, and optimization of optical systems using differentiable Fourier optics methods. The library includes a wide range of optical simulation tools, and its modular design enables easy integration of additional future developments. We believe TorchOptics can play an instrumental role in advancing future discoveries and innovations at the intersection of ``physics for machine learning'' and ``machine learning for physics.''

\section*{Declaration of competing interest}
The authors declare that they have no known competing financial interests or personal relationships that could have appeared to influence the work reported in this paper.

\section*{Acknowledgments}
This  work is supported by the European Union's Horizon 2020 research and innovation programme under the Marie Skłodowska-Curie grant agreement No.~956071. A.L.'s research is supported by the Innovate UK Smart
Grant 10043476 and the NSERC Grant EP/Y020596/1.





\bibliographystyle{elsarticle-num}







\end{document}